\documentclass[conference]{IEEEtran}
\usepackage{cite}

\ifCLASSINFOpdf
\usepackage[pdftex]{graphicx}
\graphicspath{{../pdf/}{../jpeg/}}
\DeclareGraphicsExtensions{.pdf,.jpeg,.png}
\else
\usepackage[dvips]{graphicx}
\graphicspath{{../eps/}}
\DeclareGraphicsExtensions{.eps}
\fi

\usepackage{amsmath}
\usepackage{amsfonts}
\usepackage{amssymb}
\usepackage{amsthm}
\usepackage{array}
\usepackage[caption=false,font=footnotesize]{subfig}
\usepackage{dblfloatfix}
\usepackage{tikz}
\usetikzlibrary{shapes}
\usepackage{pgfplots}
\usepackage{xcolor}
\usepackage[outline]{contour}
\contourlength{3pt}

\usepackage{multirow}
\usepackage{threeparttable}
\usepackage{hyperref}
\usepackage{xfrac}
\usepackage{ulem}
\newtheorem{thm}{Theorem}
\newtheorem{lemma}{Lemma}

\newtheorem{remark}{Remark}

\allowdisplaybreaks

\newcommand{\cc}[1]{\textcolor{black}{#1}}

\begin{document}
%
\title{On Achievable Rates of Evenly-Spaced Discrete Uniform Distributions in the IM/DD Broadcast Channel}
\author{
\IEEEauthorblockN{Zhenyu (Charlus) Zhang, Anas Chaaban}
\IEEEauthorblockA{School of Engineering, University of British Columbia, Kelowna, V1V1V7 BC, Canada\\
{zhenyu.zhang@alumni.ubc.ca}, {anas.chaaban@ubc.ca}}
}

\maketitle

\begin{abstract}
In optical wireless communications, a broadcast channel (BC) employing intensity modulation and direct detection (IM/DD) is often modeled as a peak-constrained BC. A closed-form expression for its capacity region of the peak-constrained BC is not known. This paper presents an analytical capacity inner bound for the peak-constrained Gaussian BC achieved by a class of discrete input distribution, specifically, the evenly-spaced discrete uniform distribution (ESDU). In contrast to the continuous input distribution that provides the benchmark, ESDU is more promising in the application of peak-constrained Gaussian channels. The newly obtained capacity inner bound is easily-computable and is numerically shown to be tighter than the benchmark. Besides, we remark the newly developed analytical upper bound for the ESDU rate, which is tight in all tested settings.
\end{abstract}


%
\IEEEpeerreviewmaketitle

\section{Introduction}

Optical wireless communication (OWC) has witnessed an increased research attention over the past decade, owing to the advent of LiFi \cite{haas2015lifi} in addition to emerging applications of free-space optics, visible light communications, and ultra-violet wireless communications \cite{khalighi2014survey,VLC:BeyondP2P,XuSadler}. This renewed interest in optical wireless communications led to increasing efforts in studying the capacity achieved in an OWC system using intensity modulation and direct detection (IM/DD), often modeled as a peak-constrained Gaussian channels 
with a single user \cite{Lapidoth-Moser-Wigger-2009,farid2009channel,
thangaraj2017capacity}. Driven by the growing interests in applying OWC in practice \cite{arafa2019relay,yesilkaya2022channel,tagliaferri2018nonlinear}, recent developments in this area have led to new results for peak-constrained multi-user channels, including the multiple-access channel \cite{zhou2019bounds}, \cc{the interference channel \cite{zhang2022capacity},} and the broadcast channel (BC) \cite{chaaban2016capacity}, which can model the uplink and the downlink scenarios in OWC. However, the study on the capacity of the peak-constrained multi-user channels is hindered by a major challenge. 

For a peak-constrained Gaussian channel, it is known that the capacity is achieved by a discrete input distribution \cite{smith1971information}. However, analyzing the entropy of the mixture of a Gaussian and a discrete distribution is rather challenging. As a consequence, the current literature resorts to either evaluating capacity numerically \cite{farid2009channel}, or to bounding it using continuous distributions such as uniform, truncated-exponential, or truncated Gaussian (TG) distributions leading to analytical results that exhibit a large gap to capacity \cite{Lapidoth-Moser-Wigger-2009,chaaban2016capacity} (see \cite{chaaban2021capacity} for a survey on the topic).
Moreover, although it is known that a discrete uniform distribution over an evenly-spaced alphabet (ESDU) provides a good approximation of the capacity of the peak-constrained Gaussian channel \cite{farid2009channel}, there is no closed-form expression for the rate achieved by a general ESDU (or a good rate bound thereon). Consequently, the effectiveness of ESDU in the peak-constrained Gaussian channel is only known through numerical results for specific settings \cite{farid2009channel}.
Due to the challenge in analyzing discrete channel input, the study on the capacity of the peak-constrained Gaussian BC has been hindered for a long time, as observed in the current literature. The earliest study we find in the literature on the capacity of the peak-constrained Gaussian BC is \cite{chaaban2016capacity}, which developed capacity inner bounds and outer bounds. After \cite{chaaban2016capacity}, little new knowledge about the capacity of the peak-constrained Gaussian BC appears in the literature, despite the importance of the peak-constrained Gaussian BC for OWC. 
The recent literature on the peak-constrained BC tends to accept the disadvantage of continuous input distributions in terms of rate in exchange for the advantage of having simple achievable rate expressions. For example, the recent work \cite{pham2017multi,tagliaferri2018nonlinear,arafa2019relay} are all based on analytical single-user capacity lower bounds that are derived for continuous input distributions and thus have inherited their large gap to capacity. This highlights the importance of deriving simple analytical results for the achievable rates of discrete input distributions towards developing tighter capacity bounds for the peak-constrained Gaussian BC.

In this paper, we aim to address this challenge. We propose to adopt ESDU in the peak-constrained Gaussian BC and study its achievable rate. More specifically, analytical results are derived for the peak-constrained Gaussian point-to-point (P2P) channel based on ESDU input, leading to an upper and a lower bound on its achievable rate, which are then used to obtain an analytical inner bound for the peak-constrained Gaussian BC. This ESDU-based BC inner bound is then examined numerically in comparison with the benchmark inner bound from \cite{chaaban2016capacity} which is based on a TG distribution, and is shown to achieve a larger rate region. Besides, the numerical results also show that the obtained ESDU rate upper bound is remarkably tight in all tested settings.

The rest of the paper is organized as follows. Sec. II describes the channel model and the objective. The achievable rate analysis of the ESDU and the main results of the paper are given in Sec. III. Then Sec. IV numerically examines the obtained results. Finally, Sec. V concludes this paper and introduces possible future extensions. 

\emph{Notations:} Throughout the paper, we use $\mathbb{R}_+$ to denote the set of nonnegative real numbers, $I(\cdot;\cdot)$ to denote the mutual information between two random variables, and $H(\cdot)$ and $h(\cdot)$ to denote the entropy and differential entropy of a random variable, respectively. We also use $H(p)$ to denote the binary entropy function, i.e., for $p\in[0,1]$, $H(p)=-p\log(p)-(1-p)\log(1-p)$. We use $\log(\cdot)$ to denote the base-2 logarithm, $\mathbb{P}[\cdot]$ to denote the probability of a random event, $P_X$ to denote the probability distribution of $X$, and $\mathbb{E}[\cdot]$ and $\mathbb{V}[\cdot]$ to denote the expectation and the variance of a random variable, respectively. We write $X\sim{\rm Unif}([a,b])$ to indicate that a continuous random variable, $X$, is uniformly distributed on the interval $[a,b]$, and $X\sim{\rm Unif}(\mathcal{X})$ to indicate that a discrete random variable, $X$, is uniformly distributed on the alphabet $\mathcal{X}$. Specifically, we write $X\sim\mathsf{ESDU}(A,K)$ to indicate that a discrete random variable, $X$, is uniformly distributed over the set $\{\frac{iA}{K-1}\}_{i=0}^{K-1}$, an alphabet with $K$ elements spanning $[0,A]$ with a spacing of $\frac{A}{K-1}$, or equivalently, $\mathsf{ESDU}(A,K)={\rm Unif}\bigl(\{\frac{iA}{K-1}\}_{i=0}^{K-1}\bigr)$. Finally, $X\sim\mathcal{N}(\mu,\sigma^2)$ defines a Gaussian random variable with mean $\mu$ and variance~$\sigma^2$ and $\mathsf{Q}(x)=\int_{x}^{\infty}{\frac{1}{\sqrt{2\pi}}e^{-\frac{x^2}{2}}dx}$ is the standard Gaussian tail function.

\section{Channel Model and Objective}\label{Sec:Model}
Consider a two-user OWC BC employing an IM/DD scheme. This can be modeled as a peak-constrained Gaussian BC, defined through the input output relations
\begin{align} \label{eq:bc}
	Y_i = X + Z_i, \; i=1,2,
\end{align}
where the transmitter broadcasts $X$ to receivers 1 and 2 through independent noisy transmission links, and receiver $i$ receives $Y_i=X+Z_i$, where the additive noise at the receiver is Gaussian, i.e., $Z_i\sim\mathcal{N}(0,\sigma_i^2)$. Without loss of generality (WLOG), we suppose $\sigma_1<\sigma_2$. The transmit signal satisfies nonnegativity and peak constraints so as $X\in[0,A]$. 

Using this channel, the transmitter wants to send messages $M_1$ and $M_2$ with rates $R_1$ and $R_2$ to receivers 1 and 2, respectively. Achievable rate pairs $(R_1,R_2)$ and the capacity region of this BC are defined in the standard Shannon sense, see \cite[Ch. 5]{book-NIT}. Note that this BC belongs to the family of degraded BCs, since given $\tilde{Z}_2\sim\mathcal{N}(0,\sigma_2^2-\sigma_1^2)$ and $\tilde{Y}_2=Y_1+\tilde{Z}_2$, $P_{Y_2|X}=P_{\tilde{Y}_2|X}$ and $X-Y_1-\tilde{Y}_2$ forms a Markov chain.

Thus, the capacity region of this two-user BC is the set of rate pairs $(R_1,R_2)$ that satisfy  \cite[Chap. 5]{book-NIT}
\begin{subequations}
	\label{eq:cap_bc}
	\begin{align}
		0\le R_1 &\le I(X;Y_1|U), \\
		0\le R_2 &\le I(U;Y_2),
	\end{align}
\end{subequations}
for some $P_{U,X}$, where $U$ is an auxiliary random variable that conveys the message to receiver 2. This region can be achieved through superposition coding with successive interference cancellation (SC-SIC).

The main challenge is to determine $U$ and $X$ such that the peak constraint is satisfied. The objective of this work is to provide an analytical lower bound on this capacity region when $X$ follows a discrete distribution. To this end, we propose a transmission scheme that combines SC-SIC and an ESDU, and analyze its achievable rate to obtain a closed-form expression. Details are given next.

\section{Proposed Scheme and Achievable Rate}

We propose to adopt SC-SIC in the two-user BC \eqref{eq:bc} while designing $X$ so that $X\sim\mathsf{ESDU}(A,K)$ for some $K\ge2$. The construction of $X$ is as follows. Given some integers $K_1\ge1$ and $K_2\ge1$ such that $K=K_1K_2\geq2$, define independent ESDU random variables $X_1\sim\mathsf{ESDU}\bigl(\frac{(K_1-1)A}{K-1},K_1\bigr)$ and $X_2\sim\mathsf{ESDU}\bigl(\frac{(K_2-1)K_1A}{K-1},K_2\bigr)$. The random variables $X_1$ and $X_2$ will be used to encode $M_1$ and $M_2$, respectively, using an independent and identically distributed (i.i.d.) random code, i.e., the codeword for $M_i$ is an i.i.d. sequence of realizations of $X_i$. Finally, the transmit signal is constructed by adding the codewords, and hence $X=X_1+X_2$. 

For decoding, receiver 2 decodes $M_2$ from its received signal, while receiver 1 decodes $M_2$  to obtain $X_2$, subtracts its contribution from the received signal, and finally then $M_1$. The achievable rate region is the convex hull of the union over $K_i\geq1$ of sets of rate pairs $(R_1,R_2)\in\mathbb{R}_+^2$ that satisfy
\begin{subequations}
	\label{eq:bc_rate}
	\begin{align}
		R_1 &\le I(X_1;X_1+Z_1), \label{eq:bc_rate_a}\\
		R_2 &\le I(X_2;Y_2) \\
		& = I(X;Y_2)-I(X_1;X_1+Z_2), \label{eq:bc_rate_c}
	\end{align}
\end{subequations}
where \eqref{eq:bc_rate_c} follows since $I(X_2;Y_2)=I(X_1,X_2;Y_2)-I(X_1;Y_2|X_2)=I(X;Y_2)-I(X_1;Y_2|X_2)$.

In order to simplify the evaluation of this achievable rate region, we aim to express the mutual information terms in \eqref{eq:bc_rate_a} and \eqref{eq:bc_rate_c} in closed form. To this end, we need some preliminaries, which are presented in the following subsection.

\subsection{Useful P2P Rate Bounds}
Here we present bounds on the rate that can be achieved in a peak-constrained Gaussian P2P channel using an ESDU input distribution. We start by recalling capacity bounds for the peak-constrained Gaussian P2P channel with a continuous uniform input distribution which will be useful afterwards.

\begin{lemma}[Continuous uniform distribution rate]\label{lem:cu}
	Given $X\sim{\rm Unif}([0,A])$ and $Z\sim\mathcal{N}(0,\sigma^2)$, we have $\underline{C}(A,\sigma)\leq I(X;X+Z)\leq\mathsf{E}(A,\sigma)$ where
	\begin{align}
		 \mathsf{E}(A,\sigma)\triangleq \min\Bigl\{\overline{C}(A,\sigma), \frac{1}{2}\log\Bigl(1+\frac{A^2}{12\sigma^2}\Bigr)\Bigr\}, \label{eq:E}
	\end{align}
	and $\underline{C}(A,\sigma)$ and $\overline{C}(A,\sigma)$ are the lower and upper bounds on $I(X;X+Z)$ given in \cite{Lapidoth-Moser-Wigger-2009,thangaraj2017capacity} as
	\begin{align}
		\underline{C}(A,\sigma) &= \frac{1}{2}\log\Bigl(1+\frac{A^2}{2\pi e\sigma^2}\Bigr),  \label{eq:C_underline}\\
		\overline{C}(A,\sigma) & =\min\Biggl\{\frac{1}{2}\log\Bigl(1+\frac{A^2}{4\sigma^2}\Bigr),\log\Bigl(1+\frac{A}{\sqrt{2\pi e}\sigma}\Bigr)\!\Biggr\}.\label{eq:C_overline}
	\end{align}
\end{lemma}
\begin{IEEEproof}
	The proof is given in Appendix \ref{app:lem_cu}.
\end{IEEEproof}
Note that $\underline{C}(A,\sigma)$ and $\overline{C}(A,\sigma)$ are also the lower and upper bound for the capacity of the peak-constrained Gaussian P2P channel as shown in \cite{Lapidoth-Moser-Wigger-2009,thangaraj2017capacity}.

Next, we provide lower and upper bounds on the achievable rate of an ESDU input in the following lemmas.

\begin{lemma}[ESDU rate lower bound]\label{lem:lb_esdu}
	Given $X\sim\mathsf{ESDU}(A,K)$ and $Z\sim\mathcal{N}(0,\sigma^2)$, we have $I(X;X+Z)\ge\mathsf{F}(A,K,\sigma)$ where
	\begin{align}\label{eq:F}
		\mathsf{F}(A,K,\sigma)
		= \cc{\max_{i\in\{1,2,3\}}\mathsf{F}_i(A,K,\sigma)} ,
	\end{align}
	wherein
	\begin{equation}\label{eq:F1}
		\mathsf{F}_1(A,K,\sigma) = \log(K) - H(\xi_{A,K}) - \xi_{A,K}\log(K-1),
	\end{equation}
	with $\xi_{A,K}\triangleq\frac{2(K-1)}{K}\mathsf{Q}\bigl(\frac{A}{2(K-1)\sigma}\bigr)$, and
	\begin{equation}\label{eq:F2}
		\mathsf{F}_2(A,K,\sigma) = \underline{C}\Bigl(\frac{KA}{K-1},\sigma\Bigr) - \mathsf{E}\Bigl(\frac{A}{K-1},\sigma\Bigr),
	\end{equation}
	with $\mathsf{E}$ as defined in \eqref{eq:E}, and 
	\begin{equation}\label{eq:F3}
		\cc{\mathsf{F}_3(A,K,\sigma) = -\log\biggl(\sum_{i,j\in[1,K]}\frac{\sqrt{\sfrac{e}{2}}}{K^2}e^{-\frac{(i-j)^2A^2}{4(K-1)^2\sigma^2}}\biggr)}
	\end{equation}
\end{lemma}
\begin{IEEEproof}
The proof is given in Appendix \ref{app:lem_lb_esdu}.
\end{IEEEproof}

\begin{remark}
	\cc{Regarding the three components of $\mathsf{F}(A,K,\sigma)$,} $\mathsf{F}_1(A,K,\sigma)$ is tighter than the other when $K$ is small so that $\frac{A}{K-1}$ is large, and $\mathsf{F}_2(A,K,\sigma)$ and $\mathsf{F}_3(A,K,\sigma)$ are tighter when $K$ is large so that $\frac{A}{K-1}$ is small. 
\end{remark}
As an additional remark, for an ESDU input, the new analytical lower bound $\mathsf{F}(A,K,\sigma)$ is tighter than the Ozarow-Wyner-B bound \cite{ozarow1990capacity}\cite[eq. (9)]{dytso2016interference}, which is defined as $R_{\rm OWB}(A,K,\sigma)\triangleq\log(K) - \frac{1}{2}\log\Bigl(\frac{2\pi e}{12}\Bigr) -\frac{1}{2}\log\Bigl(1+\frac{12(K-1)^2\sigma^2}{A^2}\Bigr)$. The proof is given below. Denote $\Delta=\frac{A}{K-1}$. Then, we have, 
\begin{subequations}
	\begin{align}
		&\quad \mathsf{F}(A,K,\sigma) \notag \\
		&\ge \underline{C}(K\Delta,\sigma) - \mathsf{E}(\Delta,\sigma) \\
		&\ge \frac{1}{2}\log\Bigl(1+\frac{K^2\Delta^2}{2\pi e\sigma^2}\Bigr) - \frac{1}{2}\log\Bigl(1+\frac{\Delta^2}{12\sigma^2}\Bigr) \\
		&=R_{\rm OWB}(A,K,\sigma)+\frac{1}{2}\log\Bigl(1+\frac{2\pi e\sigma^2}{K^2\Delta^2}\Bigr) \\
		&\ge R_{\rm OWB}(A,K,\sigma).
	\end{align}
\end{subequations}

\begin{lemma}[ESDU rate upper bound]\label{lem:ub_esdu}
	Given $X\sim\mathsf{ESDU}(A,K)$ and $Z\sim\mathcal{N}(0,\sigma^2)$, we have $I(X;X+Z)\le\mathsf{G}(A,K,\sigma)$ where 
	\begin{equation}\label{eq:G}
		\mathsf{G}(A,K,\sigma) = \min\bigl\{ H(K), \overline{C}(A,\sigma), \mathsf{G}'(A,K,\sigma) \bigr\}, 
	\end{equation}
	wherein
	\begin{equation}\label{eq:Gp}
		\mathsf{G}'(A,K,\sigma) = \frac{1}{2}\log\Bigl( 2^{2\mathsf{E}(\frac{KA}{K-1},\sigma)} -\frac{A^2}{2\pi e(K-1)^2\sigma^2} \Bigr),
	\end{equation}
	and $\mathsf{E}$ is defined in \eqref{eq:E}.
\end{lemma}
\begin{IEEEproof}
The proof is given in Appendix \ref{app:lem_ub_esdu}.
\end{IEEEproof}

\begin{remark}\label{rmk:esdu_ub}
	In general, $\mathsf{G}(A,K,\sigma)$ combines the bounds $H(K)$ which is a good upper bound when $K$ is small so that $\frac{A}{K-1}$ is large, $\mathsf{G}'(A,K,\sigma)$ which is good when $K$ is large so that $\frac{A}{K-1}$ is small, and $\overline{C}(A,\sigma)$ which is good overall but is a capacity upper bound (not specific for an ESDU distributed input). This conclusion is examined in Fig. \ref{fig:esdu_rate}.
\end{remark}

Now we are ready to present the main results of the paper on the achievable rate region of a peak-constrained Gaussian BC.

\subsection{BC Achievable Rate Region}
The new achievable rate region achieved using an ESDU in a peak-constrained Gaussian BC is given next.

\begin{thm}[A computable BC capacity inner bound] \label{thm:rate}
	Given a peak-constrained Gaussian BC as defined in Sec. \ref{Sec:Model}, and given any $K_i\ge1$, $i=1,2$, rate pairs $(R_1,R_2)\in\mathbb{R}_+^2$ that satisfy 	
	\begin{subequations}
		\label{eq:bc_rate_esdu}
		\begin{align}
			R_1&\le \mathsf{F}\Bigl(\frac{(K_1-1)A}{K_1K_2-1},K_1,\sigma_1\Bigr) \\*
			R_2&\le \mathsf{F}(A,K_1K_2,\sigma_2) - \mathsf{G}\Bigl(\frac{(K_1-1)A}{K_1K_2-1},K_1,\sigma_2\Bigr),
		\end{align}
	\end{subequations}
are achievable, where $\mathsf{F}$ and $\mathsf{G}$ are defined in \eqref{eq:F} and \eqref{eq:G}, respectively.
\end{thm}
\begin{IEEEproof}
The proof is obtained based on \eqref{eq:bc_rate} while using Lemmas \ref{lem:lb_esdu} and \ref{lem:ub_esdu} to lower-bound $I(X_1;X_1+Z_1)$ and $I(X;Y_2)$ and to upper-bound $I(X_1;X_1+Z_2)$, respectively, where both $X_1$ and $X$ follow the ESDU distribution.
\end{IEEEproof}

To assess the above inner bound, we use the following capacity outer bound \cite{chaaban2016capacity}.

\begin{thm}[BC capacity outer bound]\label{lem:bc_outer}
	An achievable rate pair $(R_1,R_2)\in\mathbb{R}_+^2$ in a peak-constrained Gaussian BC as defined in Sec. \ref{Sec:Model} satisfies $(R_1,R_2)\in \overline{\mathcal{G}}\triangleq\overline{\mathcal{G}}_1 \cap \overline{\mathcal{G}}_2$, where $\overline{\mathcal{G}}_1 $ is in the convex hull of the union over $\rho\in[0,1]$ of rate pairs $(R_1,R_2)\in\mathbb{R}^+$ satisfying
	\begin{subequations}
		\begin{align}
			R_1 &\le \frac{1}{2} \log\Bigl( 1+\frac{\sigma_2^2( e^{2\overline{C}(\rho A,\sigma_2)}-1 )}{\sigma_1^2} \Bigr), \\
			R_2 &\le \overline{C}(A,\sigma_2) - \overline{C}(\rho A,\sigma_2)
		\end{align}
	\end{subequations}
	and where
	\begin{equation}
		\overline{\mathcal{G}}_2 \triangleq \left\{  (R_1,R_2)\in\mathbb{R}^+ \Bigg| 
		\begin{array}{l}
			R_i\le \overline{C}(A,\sigma_i),\;i=1,2,\\
			R_1+R_2\le \overline{C}(A,\sigma_1)
		\end{array} \right\}
	\end{equation}
\end{thm}
\begin{IEEEproof}
The proof of the outer bound $\overline{\mathcal{G}}_1$ is given in \cite{chaaban2016capacity}. The proof of the outer bound $\overline{\mathcal{G}}_2$ follows due to the degradedness property \cite[Sec. 5.4]{book-NIT}, implying that the sum rate cannot not be larger than the capacity of the less noisy link (from the transmitter to receiver 1).
\end{IEEEproof}

\section{Numerical Results}\label{Sec:Numerical}
In this section, we numerically examine the obtained results. The lower and upper bounds for the ESDU rate in Lemma \ref{lem:lb_esdu} and \ref{lem:ub_esdu} are shown first, followed by the achievable rate region of the BC. \cc{Without loss of generality, we set $\sigma=\sigma_1=1$ throughout the simulations.}

\subsection{The Lower and the Upper Bounds of ESDU Rate}
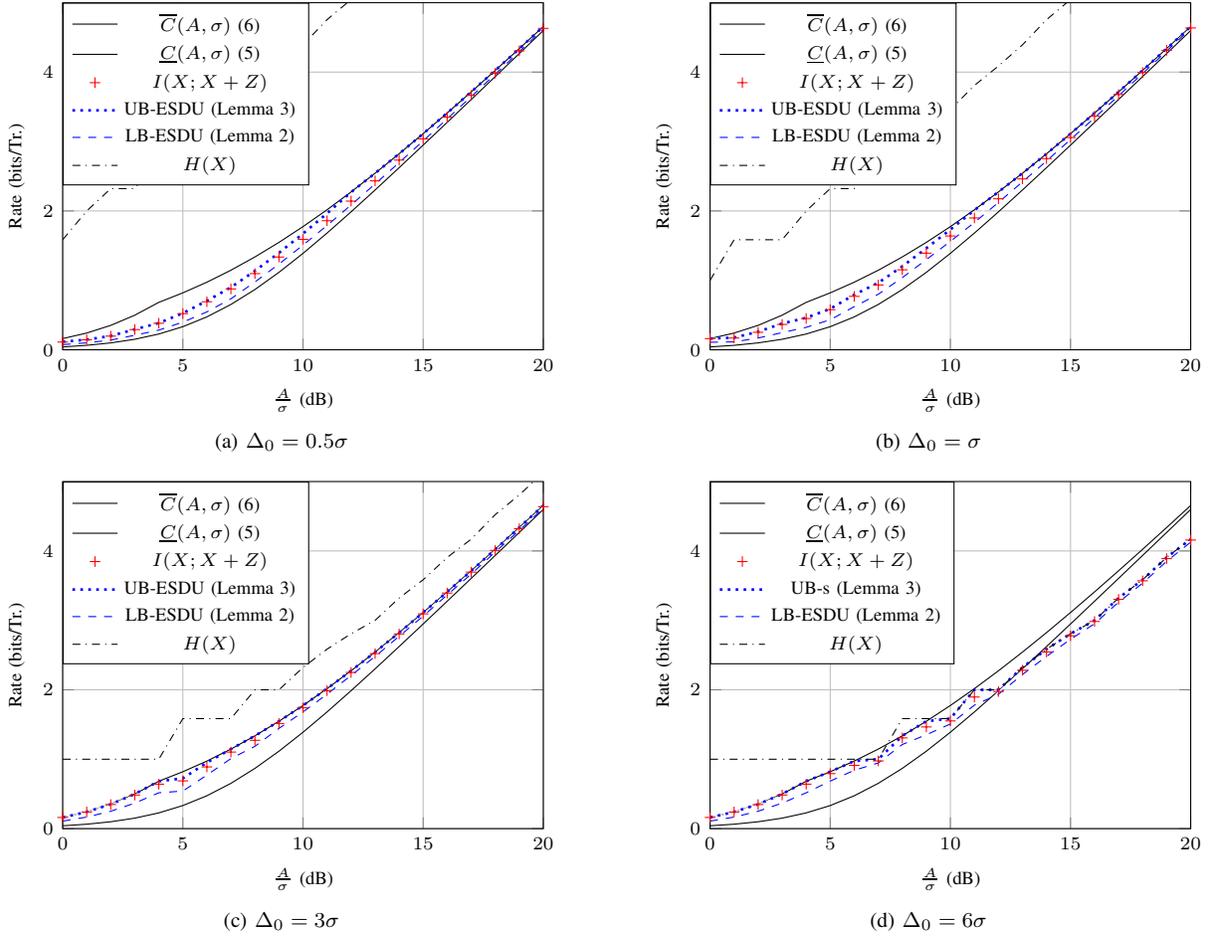
\begin{figure*}[!htbp]
	\centering
	\subfloat[$\Delta_0=0.5\sigma$]{
%
%

\begin{tikzpicture}
\begin{axis}[%
	width=.9\columnwidth,
	height=.7\columnwidth,
	xmin=0,
	xmax=20,
	xlabel={\scriptsize $\frac{A}{\sigma}$ (dB)},
	ymin=0,
	ymax=5,
	ylabel={\scriptsize Rate (bits/Tr.)},
	xmajorgrids,
	ymajorgrids,
	ticklabel style = {font=\scriptsize},
	xlabel near ticks,
	ylabel near ticks,
	legend style={ at={(axis cs: 0,5)}, anchor=north west}
]
\addplot [color=black]
table[row sep=crcr]{%
	0	0.160964047443681\\
	1	0.240764845172548\\
	2	0.351537769479087\\
	3	0.498291242926011\\
	4	0.682956368898559\\
	5	0.819814175671874\\
	6	0.973282700318478\\
	7	1.14582506250897\\
	8	1.33727268473667\\
	9	1.54697680854408\\
	10	1.77387282394493\\
	11	2.01657866458013\\
	12	2.27350918813765\\
	13	2.54298866830575\\
	14	2.82334783530117\\
	15	3.11299800861738\\
	16	3.41048059767537\\
	17	3.71449433269512\\
	18	4.02390473640102\\
	19	4.33774090510789\\
	20	4.65518421732466\\
	21	4.97555266642647\\
	22	5.29828349629194\\
	23	5.6229159120666\\
	24	5.94907492417097\\
	25	6.27645686919303\\
	26	6.60481680764183\\
	27	6.93395778595632\\
	28	7.26372183072044\\
	29	7.59398248571639\\
	30	7.92463868411843\\
};
\addlegendentry{\scriptsize $\overline{C}(A,\sigma)$ \eqref{eq:C_overline}}

\addplot [color=black]
  table[row sep=crcr]{%
0	0.0410445931048853\\
1	0.0640115453347339\\
2	0.0989770489302889\\
3	0.151139681465882\\
4	0.22678484439847\\
5	0.332468172214542\\
6	0.473534676711565\\
7	0.652461462134994\\
8	0.867958354866731\\
9	1.11544651205882\\
10	1.38857656014063\\
11	1.68085070365482\\
12	1.98667029720715\\
13	2.30171251845218\\
14	2.62287039490101\\
15	2.94801276068759\\
16	3.27572132068937\\
17	3.60507034921808\\
18	3.93546306855791\\
19	4.26651778228777\\
20	4.59799157689679\\
21	4.92973034962902\\
22	5.26163653417681\\
23	5.5936484369366\\
24	5.92572707864751\\
25	6.25784784383995\\
26	6.58999519274827\\
27	6.92215931707569\\
28	7.25433402686428\\
29	7.5865154159806\\
30	7.91870101960845\\
};
\addlegendentry{\scriptsize $\underline{C}(A,\sigma)$ \eqref{eq:C_underline}}

\addplot [color=white, mark=+, mark options={color=red}]
table[row sep=crcr]{%
	0	0.111166693415685\\
	1	0.143451091398157\\
	2	0.196791377636354\\
	3	0.290598886814533\\
	4	0.381466301931511\\
	5	0.520498650122386\\
	6	0.691748500668513\\
	7	0.876290445624623\\
	8	1.09711338235769\\
	9	1.33570294054145\\
	10	1.59082183063296\\
	11	1.85885809059391\\
	12	2.14142439211109\\
	13	2.43313500412782\\
	14	2.73233943866256\\
	15	3.03847858563571\\
	16	3.34989450951698\\
	17	3.66535355432904\\
	18	3.9842476867549\\
	19	4.3058757742271\\
	20	4.62961580676271\\
};
\addlegendentry{\scriptsize $I(X;X+Z)$}

\addplot [color=blue, dotted, line width=1pt]
table[row sep=crcr]{%
	0	0.115016970696966\\
	1	0.146089559554917\\
	2	0.199107665312887\\
	3	0.294306602358885\\
	4	0.385058219173179\\
	5	0.527072221466701\\
	6	0.704917516411174\\
	7	0.899330627245337\\
	8	1.13657765595849\\
	9	1.39552500436116\\
	10	1.67332689560707\\
	11	1.9643846464807\\
	12	2.26947827144492\\
	13	2.54298866830575\\
	14	2.82334783530117\\
	15	3.11299800861738\\
	16	3.41048059767537\\
	17	3.71449433269512\\
	18	4.02390473640102\\
	19	4.33774090510789\\
	20	4.65518421732466\\
	21	4.97555266642647\\
	22	5.29828349629194\\
	23	5.6229159120666\\
	24	5.94907492417097\\
	25	6.27645686919303\\
	26	6.60481680764183\\
	27	6.93395778595632\\
	28	7.26372183072044\\
	29	7.59398248571639\\
	30	7.92463868411843\\
};
\addlegendentry{\scriptsize UB-ESDU (Lemma \ref{lem:ub_esdu})}

\addplot [color=blue, dashed]
table[row sep=crcr]{%
0	0.0743957727703369\\
1	0.0996371576914942\\
2	0.139834501986853\\
3	0.209226120923704\\
4	0.283380386885741\\
5	0.397560090034825\\
6	0.545587162289458\\
7	0.733994892933732\\
8	0.976516225194942\\
9	1.23486219309477\\
10	1.50742090442986\\
11	1.79035892514079\\
12	2.08571887879487\\
13	2.38803497501657\\
14	2.69594664328627\\
15	3.00922555396541\\
16	3.32644232823165\\
17	3.64658295480584\\
18	3.96924947396894\\
19	4.29390793429995\\
20	4.62007408702805\\
};
\addlegendentry{\scriptsize LB-ESDU (Lemma \ref{lem:lb_esdu})}

\addplot [color=black, dash dot]
table[row sep=crcr]{%
	0	1.58496250072116\\
	1	2\\
	2	2.32192809488736\\
	3	2.32192809488736\\
	4	2.8073549220576\\
	5	3\\
	6	3.16992500144231\\
	7	3.58496250072116\\
	8	3.8073549220576\\
	9	4.08746284125034\\
	10	4.39231742277876\\
	11	4.75488750216347\\
	12	5.04439411935845\\
	13	5.35755200461808\\
	14	5.70043971814109\\
	15	6.02236781302845\\
	16	6.33985000288462\\
	17	6.6724253419715\\
	18	7\\
	19	7.32192809488736\\
	20	7.65105169117893\\
	21	7.98299357469431\\
	22	8.31288295528435\\
	23	8.64745842645492\\
	24	8.97727992349992\\
	25	9.30833903013941\\
	26	9.64024493622235\\
	27	9.97154355395077\\
	28	10.3026389237876\\
	29	10.6348110501717\\
	30	10.9665054519057\\
};
\addlegendentry{\scriptsize $H(X)$}

\end{axis}

\end{tikzpicture}%
\label{fig:newLBa}}\hspace{1cm}
	\subfloat[$\Delta_0=\sigma$]{
%
%

\begin{tikzpicture}
\begin{axis}[%
	width=.9\columnwidth,
	height=.7\columnwidth,
	xmin=0,
	xmax=20,
	xlabel={\scriptsize $\frac{A}{\sigma}$ (dB)},
	ymin=0,
	ymax=5,
	ylabel={\scriptsize Rate (bits/Tr.)},
	xmajorgrids,
	ymajorgrids,
	ticklabel style = {font=\scriptsize},
	xlabel near ticks,
	ylabel near ticks,
	legend style={ at={(axis cs: 0,5)}, anchor=north west}
]
\addplot [color=black]
table[row sep=crcr]{%
	0	0.160964047443681\\
	1	0.240764845172548\\
	2	0.351537769479087\\
	3	0.498291242926011\\
	4	0.682956368898559\\
	5	0.819814175671874\\
	6	0.973282700318478\\
	7	1.14582506250897\\
	8	1.33727268473667\\
	9	1.54697680854408\\
	10	1.77387282394493\\
	11	2.01657866458013\\
	12	2.27350918813765\\
	13	2.54298866830575\\
	14	2.82334783530117\\
	15	3.11299800861738\\
	16	3.41048059767537\\
	17	3.71449433269512\\
	18	4.02390473640102\\
	19	4.33774090510789\\
	20	4.65518421732466\\
	21	4.97555266642647\\
	22	5.29828349629194\\
	23	5.6229159120666\\
	24	5.94907492417097\\
	25	6.27645686919303\\
	26	6.60481680764183\\
	27	6.93395778595632\\
	28	7.26372183072044\\
	29	7.59398248571639\\
	30	7.92463868411843\\
};
\addlegendentry{\scriptsize $\overline{C}(A,\sigma)$ \eqref{eq:C_overline}}

\addplot [color=black]
  table[row sep=crcr]{%
0	0.0410445931048853\\
1	0.0640115453347339\\
2	0.0989770489302889\\
3	0.151139681465882\\
4	0.22678484439847\\
5	0.332468172214542\\
6	0.473534676711565\\
7	0.652461462134994\\
8	0.867958354866731\\
9	1.11544651205882\\
10	1.38857656014063\\
11	1.68085070365482\\
12	1.98667029720715\\
13	2.30171251845218\\
14	2.62287039490101\\
15	2.94801276068759\\
16	3.27572132068937\\
17	3.60507034921808\\
18	3.93546306855791\\
19	4.26651778228777\\
20	4.59799157689679\\
21	4.92973034962902\\
22	5.26163653417681\\
23	5.5936484369366\\
24	5.92572707864751\\
25	6.25784784383995\\
26	6.58999519274827\\
27	6.92215931707569\\
28	7.25433402686428\\
29	7.5865154159806\\
30	7.91870101960845\\
31	8.2508892824693\\
32	8.58307922321503\\
};
\addlegendentry{\scriptsize $\underline{C}(A,\sigma)$ \eqref{eq:C_underline}}

\addplot [color=white, mark=+, mark options={color=red}]
table[row sep=crcr]{%
	0	0.160747219796419\\
	1	0.168941695401145\\
	2	0.25164940898256\\
	3	0.364827531739841\\
	4	0.450217966394787\\
	5	0.577389155076848\\
	6	0.769635601745082\\
	7	0.933209957970678\\
	8	1.1520692437156\\
	9	1.39198470864241\\
	10	1.63918656391411\\
	11	1.89851300258094\\
	12	2.17499172546419\\
	13	2.46093438966419\\
	14	2.75400218964464\\
	15	3.05679066761853\\
	16	3.36479111358456\\
	17	3.67710363146024\\
	18	3.99373250153906\\
	19	4.31353637054518\\
	20	4.63582239098732\\
};
\addlegendentry{\scriptsize $I(X;X+Z)$}

\addplot [color=blue, dotted, line width=1pt]
table[row sep=crcr]{%
	0	0.160964047443681\\
	1	0.174664897187162\\
	2	0.260128681161297\\
	3	0.377733589216424\\
	4	0.460605597076113\\
	5	0.589911462222097\\
	6	0.793842348683285\\
	7	0.964569307668419\\
	8	1.20102948477558\\
	9	1.46270595381224\\
	10	1.73133921007734\\
	11	2.01174300054865\\
	12	2.27350918813765\\
	13	2.54298866830575\\
	14	2.82334783530117\\
	15	3.11299800861738\\
	16	3.41048059767537\\
	17	3.71449433269512\\
	18	4.02390473640102\\
	19	4.33774090510789\\
	20	4.65518421732466\\
	21	4.97555266642647\\
	22	5.29828349629194\\
	23	5.6229159120666\\
	24	5.94907492417097\\
	25	6.27645686919303\\
	26	6.60481680764183\\
	27	6.93395778595632\\
	28	7.26372183072044\\
	29	7.59398248571639\\
	30	7.92463868411843\\
};
\addlegendentry{\scriptsize UB-ESDU (Lemma \ref{lem:ub_esdu})}

\addplot [color=blue, dashed]
table[row sep=crcr]{%
	0	0.108521920877505\\
	1	0.113348256253095\\
	2	0.169411537672635\\
	3	0.246670760257236\\
	4	0.323222501215064\\
	5	0.431987854300832\\
	6	0.6211392838023\\
	7	0.799788756658652\\
	8	1.03937618414541\\
	9	1.29854374317339\\
	10	1.56144368558517\\
	11	1.83407029242138\\
	12	2.12240403879661\\
	13	2.4181827912883\\
	14	2.71925217898999\\
	15	3.02883669289851\\
	16	3.34233209064811\\
	17	3.65906607950741\\
	18	3.97929888899669\\
	19	4.30200899983856\\
	20	4.62662741225169\\
};
\addlegendentry{\scriptsize LB-ESDU (Lemma \ref{lem:lb_esdu})}

\addplot [color=black, dash dot]
table[row sep=crcr]{%
	0	1\\
	1	1.58496250072116\\
	2	1.58496250072116\\
	3	1.58496250072116\\
	4	2\\
	5	2.32192809488736\\
	6	2.32192809488736\\
	7	2.8073549220576\\
	8	3\\
	9	3.16992500144231\\
	10	3.4594316186373\\
	11	3.8073549220576\\
	12	4.08746284125034\\
	13	4.39231742277876\\
	14	4.75488750216347\\
	15	5.04439411935845\\
	16	5.35755200461808\\
	17	5.70043971814109\\
	18	6.02236781302845\\
	19	6.33985000288462\\
	20	6.65821148275179\\
	21	6.98868468677217\\
	22	7.32192809488736\\
	23	7.65105169117893\\
	24	7.98299357469431\\
	25	8.31288295528435\\
	26	8.64385618977473\\
	27	8.97441458980553\\
	28	9.3037807481771\\
	29	9.63662462054365\\
	30	9.96722625883599\\
};
\addlegendentry{\scriptsize $H(X)$}

\end{axis}

\end{tikzpicture}%
\label{fig:newLBb}}
	
	\subfloat[$\Delta_0=3\sigma$]{
%
%

\begin{tikzpicture}
\begin{axis}[%
	width=.9\columnwidth,
	height=.7\columnwidth,
	xmin=0,
	xmax=20,
	xlabel={\scriptsize $\frac{A}{\sigma}$ (dB)},
	ymin=0,
	ymax=5,
	ylabel={\scriptsize Rate (bits/Tr.)},
	xmajorgrids,
	ymajorgrids,
	ticklabel style = {font=\scriptsize},
	xlabel near ticks,
	ylabel near ticks,
	legend style={ at={(axis cs: 0,5)}, anchor=north west}
]
\addplot [color=black]
table[row sep=crcr]{%
	0	0.160964047443681\\
	1	0.240764845172548\\
	2	0.351537769479087\\
	3	0.498291242926011\\
	4	0.682956368898559\\
	5	0.819814175671874\\
	6	0.973282700318478\\
	7	1.14582506250897\\
	8	1.33727268473667\\
	9	1.54697680854408\\
	10	1.77387282394493\\
	11	2.01657866458013\\
	12	2.27350918813765\\
	13	2.54298866830575\\
	14	2.82334783530117\\
	15	3.11299800861738\\
	16	3.41048059767537\\
	17	3.71449433269512\\
	18	4.02390473640102\\
	19	4.33774090510789\\
	20	4.65518421732466\\
	21	4.97555266642647\\
	22	5.29828349629194\\
	23	5.6229159120666\\
	24	5.94907492417097\\
	25	6.27645686919303\\
	26	6.60481680764183\\
	27	6.93395778595632\\
	28	7.26372183072044\\
	29	7.59398248571639\\
	30	7.92463868411843\\
};
\addlegendentry{\scriptsize $\overline{C}(A,\sigma)$ \eqref{eq:C_overline}}

\addplot [color=black]
  table[row sep=crcr]{%
0	0.0410445931048853\\
1	0.0640115453347339\\
2	0.0989770489302889\\
3	0.151139681465882\\
4	0.22678484439847\\
5	0.332468172214542\\
6	0.473534676711565\\
7	0.652461462134994\\
8	0.867958354866731\\
9	1.11544651205882\\
10	1.38857656014063\\
11	1.68085070365482\\
12	1.98667029720715\\
13	2.30171251845218\\
14	2.62287039490101\\
15	2.94801276068759\\
16	3.27572132068937\\
17	3.60507034921808\\
18	3.93546306855791\\
19	4.26651778228777\\
20	4.59799157689679\\
21	4.92973034962902\\
22	5.26163653417681\\
23	5.5936484369366\\
24	5.92572707864751\\
25	6.25784784383995\\
26	6.58999519274827\\
27	6.92215931707569\\
28	7.25433402686428\\
29	7.5865154159806\\
30	7.91870101960845\\
31	8.2508892824693\\
32	8.58307922321503\\
};
\addlegendentry{\scriptsize $\underline{C}(A,\sigma)$ \eqref{eq:C_underline}}

\addplot [color=white, mark=+, mark options={color=red}]
table[row sep=crcr]{%
	0	0.160747219796419\\
	1	0.239789399435927\\
	2	0.347611259200077\\
	3	0.484406899822186\\
	4	0.640509531920561\\
	5	0.688015295559687\\
	6	0.888912585413278\\
	7	1.10273175617198\\
	8	1.27217285586005\\
	9	1.51464137896551\\
	10	1.74015507256844\\
	11	1.98717577010045\\
	12	2.24969559914613\\
	13	2.52126410399551\\
	14	2.80409104593697\\
	15	3.093805507367\\
	16	3.39326249787524\\
	17	3.69513804296769\\
	18	4.00802735277279\\
	19	4.31994317801489\\
	20	4.63792338498693\\
};
\addlegendentry{\scriptsize $I(X;X+Z)$}

\addplot [color=blue, dotted, line width=1pt]
table[row sep=crcr]{%
	0	0.160964047443681\\
	1	0.240764845172548\\
	2	0.351537769479087\\
	3	0.498291242926011\\
	4	0.682956368898559\\
	5	0.724087179335407\\
	6	0.951458697944192\\
	7	1.14582506250897\\
	8	1.33727268473667\\
	9	1.54697680854408\\
	10	1.77387282394493\\
	11	2.01657866458013\\
	12	2.27350918813765\\
	13	2.54298866830575\\
	14	2.82334783530117\\
	15	3.11299800861738\\
	16	3.41048059767537\\
	17	3.71449433269512\\
	18	4.02390473640102\\
	19	4.33774090510789\\
	20	4.65518421732466\\
	21	4.97555266642647\\
	22	5.29828349629194\\
	23	5.6229159120666\\
	24	5.94907492417097\\
	25	6.27645686919303\\
	26	6.60481680764183\\
	27	6.93395778595632\\
	28	7.26372183072044\\
	29	7.59398248571639\\
	30	7.92463868411843\\
};
\addlegendentry{\scriptsize UB-ESDU (Lemma \ref{lem:ub_esdu})}

\addplot [color=blue, dashed]
table[row sep=crcr]{%
	0	0.108521920877505\\
	1	0.1665046422357\\
	2	0.25084652069915\\
	3	0.367537820554469\\
	4	0.516688356442974\\
	5	0.541166068344102\\
	6	0.771132121940342\\
	7	1.00897939972481\\
	8	1.18562421899499\\
	9	1.44423183645674\\
	10	1.68083094152384\\
	11	1.9371138445676\\
	12	2.20648184712689\\
	13	2.47997637032356\\
	14	2.77011085417747\\
	15	3.06182819782385\\
	16	3.36618628342646\\
	17	3.66721711511226\\
	18	3.98582728929325\\
	19	4.2968836943589\\
	20	4.61656550026481\\
};
\addlegendentry{\scriptsize LB-ESDU (Lemma \ref{lem:lb_esdu})}

\addplot [color=black, dash dot]
table[row sep=crcr]{%
	0	1\\
	1	1\\
	2	1\\
	3	1\\
	4	1\\
	5	1.58496250072116\\
	6	1.58496250072116\\
	7	1.58496250072116\\
	8	2\\
	9	2\\
	10	2.32192809488736\\
	11	2.58496250072116\\
	12	2.8073549220576\\
	13	3\\
	14	3.32192809488736\\
	15	3.58496250072116\\
	16	3.90689059560852\\
	17	4.16992500144231\\
	18	4.52356195605701\\
	19	4.8073549220576\\
	20	5.12928301694497\\
	21	5.4262647547021\\
	22	5.75488750216347\\
	23	6.08746284125034\\
	24	6.4093909361377\\
	25	6.74146698640115\\
	26	7.06608919045777\\
	27	7.40087943628218\\
	28	7.7279204545632\\
	29	8.05528243550119\\
	30	8.38801728534514\\
};
\addlegendentry{\scriptsize $H(X)$}

\end{axis}

\end{tikzpicture}%
\label{fig:newLBc}}\hspace{1cm}
	\subfloat[$\Delta_0=6\sigma$]{
%
%

\begin{tikzpicture}
\begin{axis}[%
	width=.9\columnwidth,
	height=.7\columnwidth,
	xmin=0,
	xmax=20,
	xlabel={\scriptsize $\frac{A}{\sigma}$ (dB)},
	ymin=0,
	ymax=5,
	ylabel={\scriptsize Rate (bits/Tr.)},
	xmajorgrids,
	ymajorgrids,
	ticklabel style = {font=\scriptsize},
	xlabel near ticks,
	ylabel near ticks,
	legend style={ at={(axis cs: 0,5)}, anchor=north west}
]
\addplot [color=black]
table[row sep=crcr]{%
	0	0.160964047443681\\
	1	0.240764845172548\\
	2	0.351537769479087\\
	3	0.498291242926011\\
	4	0.682956368898559\\
	5	0.819814175671874\\
	6	0.973282700318478\\
	7	1.14582506250897\\
	8	1.33727268473667\\
	9	1.54697680854408\\
	10	1.77387282394493\\
	11	2.01657866458013\\
	12	2.27350918813765\\
	13	2.54298866830575\\
	14	2.82334783530117\\
	15	3.11299800861738\\
	16	3.41048059767537\\
	17	3.71449433269512\\
	18	4.02390473640102\\
	19	4.33774090510789\\
	20	4.65518421732466\\
	21	4.97555266642647\\
	22	5.29828349629194\\
	23	5.6229159120666\\
	24	5.94907492417097\\
	25	6.27645686919303\\
	26	6.60481680764183\\
	27	6.93395778595632\\
	28	7.26372183072044\\
	29	7.59398248571639\\
	30	7.92463868411843\\
};
\addlegendentry{\scriptsize $\overline{C}(A,\sigma)$ \eqref{eq:C_overline}}

\addplot [color=black]
  table[row sep=crcr]{%
0	0.0410445931048853\\
1	0.0640115453347339\\
2	0.0989770489302889\\
3	0.151139681465882\\
4	0.22678484439847\\
5	0.332468172214542\\
6	0.473534676711565\\
7	0.652461462134994\\
8	0.867958354866731\\
9	1.11544651205882\\
10	1.38857656014063\\
11	1.68085070365482\\
12	1.98667029720715\\
13	2.30171251845218\\
14	2.62287039490101\\
15	2.94801276068759\\
16	3.27572132068937\\
17	3.60507034921808\\
18	3.93546306855791\\
19	4.26651778228777\\
20	4.59799157689679\\
21	4.92973034962902\\
22	5.26163653417681\\
23	5.5936484369366\\
24	5.92572707864751\\
25	6.25784784383995\\
26	6.58999519274827\\
27	6.92215931707569\\
28	7.25433402686428\\
29	7.5865154159806\\
30	7.91870101960845\\
31	8.2508892824693\\
32	8.58307922321503\\
};
\addlegendentry{\scriptsize $\underline{C}(A,\sigma)$ \eqref{eq:C_underline}}

\addplot [color=white, mark=+, mark options={color=red}]
table[row sep=crcr]{%
	0	0.160747219796419\\
	1	0.239789399435927\\
	2	0.347611259200077\\
	3	0.484406899822186\\
	4	0.640509531920561\\
	5	0.792911452667765\\
	6	0.910932652831003\\
	7	0.975581883071134\\
	8	1.3068964620297\\
	9	1.46493428565391\\
	10	1.55186783979888\\
	11	1.89585819003308\\
	12	1.97502515506517\\
	13	2.28156219085441\\
	14	2.54492912468225\\
	15	2.77829191381679\\
	16	2.98407077597365\\
	17	3.30229132048401\\
	18	3.56948744197791\\
	19	3.88954304926908\\
	20	4.15725772303727\\
};
\addlegendentry{\scriptsize $I(X;X+Z)$}

\addplot [color=blue, dotted, line width=1pt]
table[row sep=crcr]{%
	0	0.160964047443681\\
	1	0.240764845172548\\
	2	0.351537769479087\\
	3	0.498291242926011\\
	4	0.682956368898559\\
	5	0.819814175671874\\
	6	0.973282700318478\\
	7	1\\
	8	1.33727268473667\\
	9	1.54697680854408\\
	10	1.58496250072116\\
	11	2\\
	12	2\\
	13	2.32192809488736\\
	14	2.58496250072116\\
	15	2.8073549220576\\
	16	3\\
	17	3.32192809488736\\
	18	3.58496250072116\\
	19	3.90689059560852\\
	20	4.16992500144231\\
	21	4.4594316186373\\
	22	4.8073549220576\\
	23	5.12928301694497\\
	24	5.4262647547021\\
	25	5.75488750216347\\
	26	6.08746284125034\\
	27	6.4093909361377\\
	28	6.74146698640115\\
	29	7.06608919045777\\
	30	7.39231742277876\\
};
\addlegendentry{\scriptsize UB-s (Lemma \ref{lem:ub_esdu})}

\addplot [color=blue, dashed]
table[row sep=crcr]{%
	0	0.108521920877505\\
	1	0.1665046422357\\
	2	0.25084652069915\\
	3	0.367537820554469\\
	4	0.516688356442974\\
	5	0.684892100400982\\
	6	0.840595179737346\\
	7	0.94630200824637\\
	8	1.21204075535104\\
	9	1.35238295939742\\
	10	1.50752435326455\\
	11	1.77866408945475\\
	12	1.93587144164999\\
	13	2.2202487470282\\
	14	2.48088464790845\\
	15	2.72707879149402\\
	16	2.9521961436417\\
	17	3.26253975299423\\
	18	3.53594690949878\\
	19	3.85108617493521\\
	20	4.1271031730895\\
};
\addlegendentry{\scriptsize LB-ESDU (Lemma \ref{lem:lb_esdu})}

\addplot [color=black, dash dot]
table[row sep=crcr]{%
	0	1\\
	1	1\\
	2	1\\
	3	1\\
	4	1\\
	5	1\\
	6	1\\
	7	1\\
	8	1.58496250072116\\
	9	1.58496250072116\\
	10	1.58496250072116\\
	11	2\\
	12	2\\
	13	2.32192809488736\\
	14	2.58496250072116\\
	15	2.8073549220576\\
	16	3\\
	17	3.32192809488736\\
	18	3.58496250072116\\
	19	3.90689059560852\\
	20	4.16992500144231\\
	21	4.4594316186373\\
	22	4.8073549220576\\
	23	5.12928301694497\\
	24	5.4262647547021\\
	25	5.75488750216347\\
	26	6.08746284125034\\
	27	6.4093909361377\\
	28	6.74146698640115\\
	29	7.06608919045777\\
	30	7.39231742277876\\
};
\addlegendentry{\scriptsize $H(X)$}

\end{axis}

\end{tikzpicture}%
\label{fig:newLBd}}
	\caption{The lower and upper bounds on the rate $I(X;X+Z)$ achieved by an ESDU input distribution in a peak-constrained Gaussian P2P channel.}
	\label{fig:esdu_rate}
\end{figure*}

Let $X\sim\mathsf{ESDU}(A,K)$ be the input of a peak-constrained Gaussian P2P channel with a peak constraint $A$, and with output $X+Z$ where $Z\sim\mathcal{N}(0,\sigma^2)$. The achievable rate $I(X;X+Z)$ is lower- and upper-bounded as in Lemmas \ref{lem:lb_esdu} and \ref{lem:ub_esdu}, respectively. To plot these bounds, we let $K=\max\bigl\{2,\lceil\frac{A}{\Delta_0}\rceil+1\bigr\}$ and we consider $\Delta_0=0.5i\sigma$, $i=1,\dots,20$ in our simulation. Fig. \ref{fig:esdu_rate} shows the comparison between the lower bound, the upper bound, and $I(X;X+Z)$, under some representative settings. We also plot $\overline{C}(A,\sigma)$ and $\underline{C}(A,\sigma)$ as a benchmark, and plot $H(X)$ to examine Remark \ref{rmk:esdu_ub}. It can be seen in Fig. \ref{fig:esdu_rate} that $I(X;X+Z)$ always lies in between the obtained ESDU bounds, where the upper bound remains very close to $I(X;X+Z)$ in all tested settings.

\subsection{ESDU-based BC Capacity Bounds}
\begin{figure*}[!htbp]
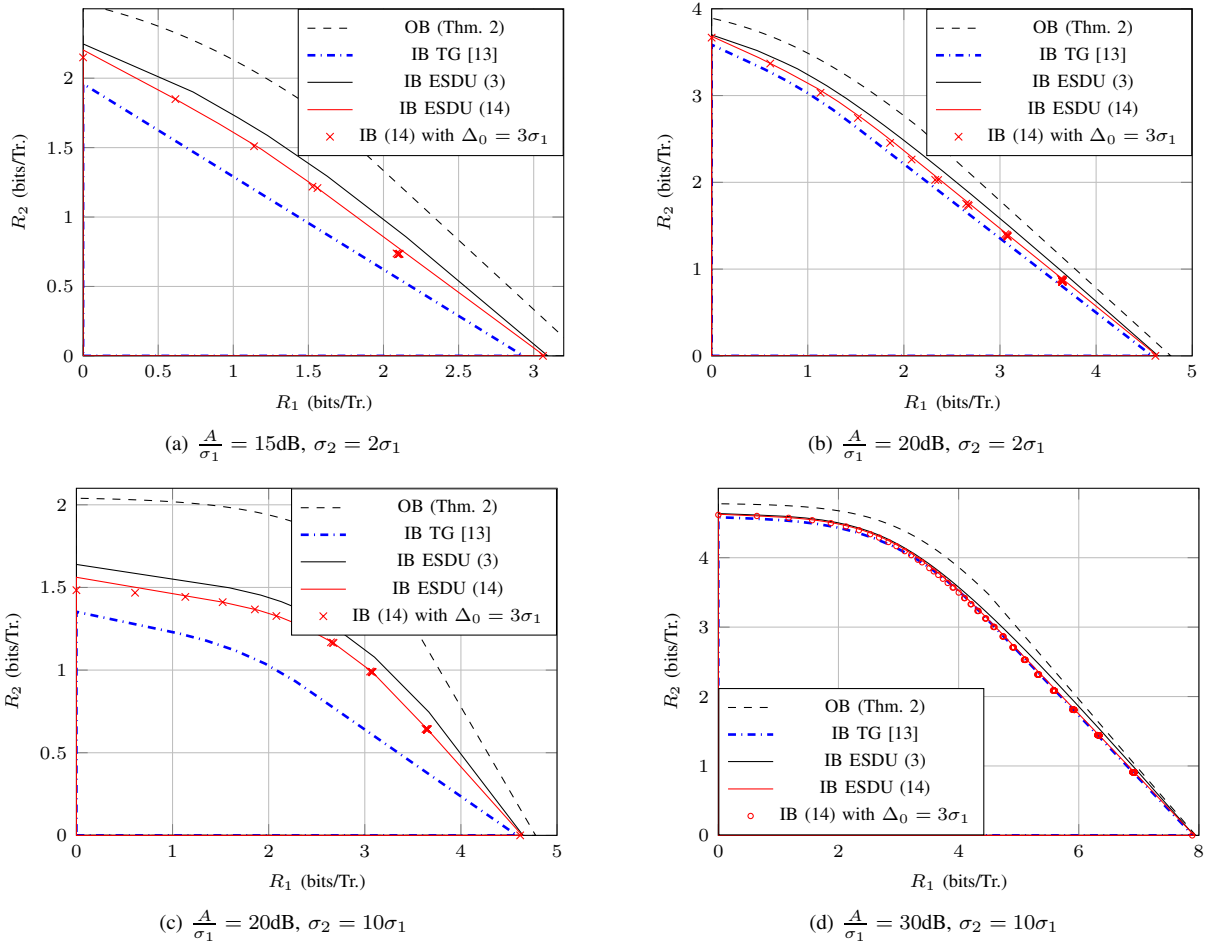

	\centering
	\subfloat[$\frac{A}{\sigma_1}=15$dB, $\sigma_2=2\sigma_1$]{
%
%
\definecolor{mycolor1}{rgb}{0.00000,0.45098,0.74118}%
\begin{tikzpicture}

\begin{axis}[%
	width=.9\columnwidth,
	height=.7\columnwidth,
	xmin=0,
	xmax=3.2,
	xlabel={\scriptsize $R_1$ (bits/Tr.)},
	ymin=0,
	ymax=2.5,
	ylabel={\scriptsize $R_2$ (bits/Tr.)},
	ytick={0,.5,1,1.5,2},
	xmajorgrids,
	ymajorgrids,
	ticklabel style = {font=\scriptsize},
	xlabel near ticks,
	ylabel near ticks,
	legend style={ at={(axis cs: 3.2,2.5)}, anchor=north east}
]
\addplot [color=black, dashed]
  table[row sep=crcr]{%
0		2.53579697	\\
0		0	\\
3.33414258		0	\\
1.608428161		1.725645299	\\
1.601338551		1.730998961	\\
1.594218214		1.736362825	\\
1.58706693		1.741736822	\\
1.57988448		1.74712088	\\
1.57267064		1.752514923	\\
1.565425189		1.757918872	\\
1.558147899		1.763332646	\\
1.550838545		1.76875616	\\
1.543496901		1.774189326	\\
1.536122734		1.779632053	\\
1.528715818		1.785084247	\\
1.521275919		1.79054581	\\
1.513802805		1.79601664	\\
1.506296243		1.801496632	\\
1.498755996		1.806985678	\\
1.49118183		1.812483665	\\
1.483573506		1.817990477	\\
1.475930787		1.823505993	\\
1.468253434		1.82903009	\\
1.460541207		1.834562639	\\
1.452793864		1.840103508	\\
1.445011165		1.84565256	\\
1.437192866		1.851209654	\\
1.429338725		1.856774643	\\
1.421448497		1.862347379	\\
1.413521941		1.867927706	\\
1.40555881		1.873515464	\\
1.397558858		1.879110489	\\
1.389521842		1.884712612	\\
1.381447516		1.890321658	\\
1.373335636		1.895937447	\\
1.365185956		1.901559794	\\
1.356998228		1.907188509	\\
1.348772211		1.912823395	\\
1.34050766		1.918464252	\\
1.332204328		1.924110873	\\
1.323861977		1.929763042	\\
1.315480361		1.935420543	\\
1.30705924		1.941083149	\\
1.298598373		1.946750629	\\
1.290097522		1.952422746	\\
1.28155645		1.958099254	\\
1.272974922		1.963779904	\\
1.264352703		1.969464438	\\
1.255689563		1.975152592	\\
1.246985273		1.980844093	\\
1.238239605		1.986538665	\\
1.229452336		1.992236021	\\
1.220623248		1.997935869	\\
1.211752121		2.003637907	\\
1.202838744		2.009341829	\\
1.193882907		2.015047319	\\
1.184884406		2.020754052	\\
1.175843039		2.026461698	\\
1.166758612		2.032169916	\\
1.157630935		2.037878359	\\
1.148459825		2.043586671	\\
1.139245106		2.049294486	\\
1.129986603		2.055001431	\\
1.120684156		2.060707124	\\
1.111337606		2.066411172	\\
1.101946806		2.072113177	\\
1.092511618		2.077812728	\\
1.083031912		2.083509406	\\
1.073507564		2.089202783	\\
1.063938467		2.094892421	\\
1.054324522		2.100577872	\\
1.044665641		2.106258679	\\
1.03496175		2.111934374	\\
1.025212786		2.11760448	\\
1.015418701		2.123268508	\\
1.005579465		2.128925961	\\
0.995695058		2.134576329	\\
0.98576548		2.140219094	\\
0.975790745		2.145853726	\\
0.965770891		2.151479683	\\
0.955705971		2.157096415	\\
0.945596059		2.162703358	\\
0.935441253		2.16829994	\\
0.925241669		2.173885575	\\
0.91499745		2.179459667	\\
0.904708768		2.185021609	\\
0.894375813		2.190570782	\\
0.88399881		2.196106555	\\
0.873578011		2.201628287	\\
0.863113696		2.207135324	\\
0.852606182		2.212627002	\\
0.842055819		2.218102643	\\
0.831462991		2.223561559	\\
0.820828121		2.22900305	\\
0.810151672		2.234426403	\\
0.799434146		2.239830895	\\
0.78867609		2.245215791	\\
0.777878099		2.250580343	\\
0.76704081		2.255923791	\\
0.756164917		2.261245367	\\
0.745251159		2.266544286	\\
0.734300338		2.271819756	\\
0.723313305		2.27707097	\\
0.712290976		2.282297113	\\
0.70123433		2.287497355	\\
0.69014441		2.292670857	\\
0.679022326		2.297816769	\\
0.667869265		2.302934229	\\
0.656686481		2.308022365	\\
0.645475313		2.313080294	\\
0.634237177		2.318107123	\\
0.622973575		2.323101947	\\
0.611686098		2.328063854	\\
0.600376428		2.332991919	\\
0.58904634		2.33788521	\\
0.577697716		2.342742783	\\
0.566332532		2.347563689	\\
0.554952877		2.352346966	\\
0.543560949		2.357091646	\\
0.532159063		2.361796753	\\
0.520749652		2.366461303	\\
0.509335272		2.371084304	\\
0.49791861		2.375664758	\\
0.486502484		2.38020166	\\
0.475089847		2.384694	\\
0.463683793		2.389140761	\\
0.452287564		2.393540923	\\
0.440904549		2.397893459	\\
0.42953829		2.402197339	\\
0.418192489		2.406451531	\\
0.406871009		2.410654997	\\
0.395577875		2.414806701	\\
0.384317287		2.4189056	\\
0.373093612		2.422950655	\\
0.361911395		2.426940822	\\
0.350775362		2.430875062	\\
0.339690416		2.434752333	\\
0.328661646		2.438571596	\\
0.317694327		2.442331815	\\
0.30679392		2.446031956	\\
0.295966074		2.44967099	\\
0.285216627		2.45324789	\\
0.274551604		2.456761639	\\
0.263977215		2.460211221	\\
0.253499857		2.463595631	\\
0.243126113		2.466913869	\\
0.23286274		2.470164946	\\
0.222716676		2.47334788	\\
0.212695022		2.476461702	\\
0.202805051		2.479505452	\\
0.193054186		2.482478183	\\
0.183450004		2.485378961	\\
0.174000217		2.488206866	\\
0.164712666		2.490960991	\\
0.155595309		2.493640446	\\
0.146656209		2.496244358	\\
0.137903511		2.49877187	\\
0.12934544		2.501222143	\\
0.120990269		2.503594358	\\
0.112846315		2.505887716	\\
0.104921908		2.508101436	\\
0.097225374		2.510234763	\\
0.089765013		2.51228696	\\
0.08254908		2.514257316	\\
0.075585749		2.516145142	\\
0.068883101		2.517949775	\\
0.06244909		2.519670578	\\
0.056291518		2.521306938	\\
0.050418009		2.522858269	\\
0.04483598		2.524324016	\\
0.039552611		2.525703647	\\
0.034574827		2.526996663	\\
0.029909254		2.528202593	\\
0.025562207		2.529320995	\\
0.021539655		2.530351459	\\
0.017847195		2.531293604	\\
0.014490032		2.532147083	\\
0.011472954		2.532911579	\\
0.008800307		2.533586808	\\
0.006475975		2.534172518	\\
0.004503365		2.53466849	\\
0.002885392		2.53507454	\\
0.001624453		2.535390514	\\
0.000722431		2.535616295	\\
0.000180676		2.535751797	\\
0		2.53579697	\\
};
\addlegendentry{\scriptsize OB (Thm. \ref{lem:bc_outer})}

\addplot [color=blue, dashdotted, line width=1.0pt]
  table[row sep=crcr]{%
0		0.0000000000	\\
2.930692136		0.0000000000	\\
0		1.9593052756	\\
0		1.9592698630	\\
0		1.9592252800	\\
0		1.9591691515	\\
0		1.9590984869	\\
0		1.9590095208	\\
0		1.9588975123	\\
0		1.9587564889	\\
0		1.9585789330	\\
0		1.9583553739	\\
0		1.9580738828	\\
0		1.9577194377	\\
0		1.9572730964	\\
0		1.9567109869	\\
0		1.9560030477	\\
0		1.9551113357	\\
0		1.9539879918	\\
0		1.9525726034	\\
0		1.9507888579	\\
0		1.9485402739	\\
0		1.9457047359	\\
0		1.9421274688	\\
0		0.0000000000	\\
};
\addlegendentry{\scriptsize IB TG \cite{chaaban2016capacity}}

\addplot [color=black]
  table[row sep=crcr]{%
0	0\\
2.77829191381679	0\\
2.91793010273406	0\\
3.00646437009226	0\\
3.05679066761853	0\\
3.08703652445301	0\\
3.09380550736701	0\\
2.1508199196887	0.856663027159621\\
1.62552130319156	1.29405989329123\\
1.22784419773721	1.58746176409012\\
1.03149738282286	1.71558793174077\\
0.732240818641114	1.90067863707386\\
0	2.24688289093769\\
0	2.23767541306665\\
0	2.22624180466905\\
0	2.1988578354597\\
0	2.17195349879009\\
0	2.13838534331819\\
0	0\\
};
\addlegendentry{\scriptsize IB ESDU \eqref{eq:bc_rate} }

\addplot [color=red]
  table[row sep=crcr]{%
0	0\\
2.72707879149402	0\\
2.79477416235633	0\\
2.89788541078709	0\\
2.99257038902128	0\\
3.00922555396541	0\\
3.02883669289851	0\\
3.06182819782385	0\\
3.06317741440918	0\\
3.07233992121169	0\\
1.56038369189476	1.2086163461867\\
1.18835629058238	1.48744863879417\\
0.902672423851222	1.67301146165802\\
0.614593593172419	1.84936562997861\\
0	2.20368120756935\\
0	2.19571678555271\\
0	2.18218428908362\\
0	2.14975213776292\\
0	2.13507729058228\\
0	2.11923860191378\\
0	2.08255381235176\\
0	2.06299837716425\\
0	2.04679680495197\\
0	0\\
};
\addlegendentry{\scriptsize IB ESDU \eqref{eq:bc_rate_esdu}}

\addplot [color=white,draw=none,  mark size=2.0pt, mark=x, mark options={solid, red}]
table[row sep=crcr]{%
0	2.14975213776292\\
0.614593593172419	1.84936562997861\\
1.1382485113329	1.50968492441925\\
1.52736018075666	1.2190026617744\\
1.56038369189476	1.2086163461867\\
2.08685453687196	0.738144815193142\\
2.10640221775589	0.736174247575017\\
2.10621813946571	0.734690256056108\\
2.10223803717872	0.733548649820954\\
2.09799817161639	0.732649137263544\\
2.09410864706785	0.731924750577867\\
3.06182819782385	0\\
};
\addlegendentry{\scriptsize  IB \eqref{eq:bc_rate_esdu} with $\Delta_0=3\sigma_1$}

\end{axis}

\end{tikzpicture}%
\label{fig:bca}}\hspace{1cm}
	\subfloat[$\frac{A}{\sigma_1}=20$dB, $\sigma_2=2\sigma_1$]{
%
%
\definecolor{mycolor1}{rgb}{0.00000,0.45098,0.74118}%
\begin{tikzpicture}

\begin{axis}[%
	width=.9\columnwidth,
	height=.7\columnwidth,
	xmin=0,
	xmax=5,
	xlabel={\scriptsize $R_1$ (bits/Tr.)},
	ymin=0,
	ymax=4,
	ylabel={\scriptsize $R_2$ (bits/Tr.)},
	xmajorgrids,
	ymajorgrids,
	ticklabel style = {font=\scriptsize},
	xlabel near ticks,
	ylabel near ticks,
	legend style={ at={(axis cs: 5,4)}, anchor=north east}
]
\addplot [color=black, dashed]
  table[row sep=crcr]{%
0		3.891775125	\\
0		0	\\
4.777117572		0	\\
2.171285164		2.621010917	\\
2.15953443		2.631237132	\\
2.143700521		2.644982341	\\
2.127700485		2.658831003	\\
2.111531031		2.672783999	\\
2.095188779		2.686842186	\\
2.078670254		2.701006401	\\
2.061971887		2.71527745	\\
2.045090011		2.729656113	\\
2.028020856		2.744143134	\\
2.010760548		2.758739219	\\
1.993305108		2.773445032	\\
1.975650442		2.788261193	\\
1.957792346		2.803188267	\\
1.939726495		2.818226764	\\
1.921448447		2.833377131	\\
1.902953634		2.848639744	\\
1.884237363		2.864014906	\\
1.865294808		2.879502836	\\
1.846121012		2.895103662	\\
1.82671088		2.910817413	\\
1.80705918		2.92664401	\\
1.787160535		2.942583255	\\
1.767009425		2.958634823	\\
1.746600181		2.97479825	\\
1.725926986		2.991072919	\\
1.704983873		3.00745805	\\
1.683764719		3.023952684	\\
1.662263253		3.04055567	\\
1.640473047		3.057265647	\\
1.618387524		3.074081031	\\
1.595999955		3.090999992	\\
1.573303464		3.108020438	\\
1.550291033		3.125139995	\\
1.526955505		3.142355984	\\
1.503289596		3.159665396	\\
1.479285898		3.177064871	\\
1.454936899		3.19455067	\\
1.430234994		3.212118647	\\
1.405172504		3.229764222	\\
1.379741699		3.247482349	\\
1.353934826		3.265267483	\\
1.32774414		3.283113549	\\
1.301161947		3.301013901	\\
1.274180646		3.318961295	\\
1.246792784		3.336947842	\\
1.218991126		3.354964972	\\
1.190768724		3.373003394	\\
1.162119012		3.391053056	\\
1.133035907		3.409103098	\\
1.103513932		3.427141816	\\
1.073548361		3.445156615	\\
1.043135381		3.46313397	\\
1.012272289		3.481059383	\\
0.980957712		3.498917343	\\
0.94919187		3.516691292	\\
0.916976875		3.534363589	\\
0.884317075		3.551915479	\\
0.851219455		3.569327067	\\
0.817694094		3.586577302	\\
0.783754687		3.60364396	\\
0.749419141		3.620503643	\\
0.714710254		3.637131785	\\
0.679656478		3.653502664	\\
0.644292776		3.669589437	\\
0.608661576		3.685364182	\\
0.57281383		3.700797954	\\
0.536810153		3.715860862	\\
0.500722069		3.730522163	\\
0.464633309		3.744750375	\\
0.428641155		3.758513405	\\
0.392857782		3.77177871	\\
0.357411536		3.784513459	\\
0.322448058		3.796684734	\\
0.288131165		3.80825974	\\
0.25464334		3.819206033	\\
0.222185688		3.829491767	\\
0.190977171		3.839085951	\\
0.161252962		3.847958718	\\
0.13326172		3.856081597	\\
0.107261666		3.863427791	\\
0.083515385		3.869972445	\\
0.062283358		3.87569291	\\
0.043816407		3.88056899	\\
0.028347334		3.884583173	\\
0.016082215		3.887720833	\\
0.007191952		3.889970404	\\
0.001804721		3.891323521	\\
0		3.891775125	\\
};
\addlegendentry{\scriptsize OB (Thm. \ref{lem:bc_outer})}

\addplot [color=blue, dashdotted, line width=1.0pt]
  table[row sep=crcr]{%
0		0	\\
4.582802923		0	\\
1.52124385		2.621694466	\\
1.258019052		2.839676526	\\
1.004656391		3.025136502	\\
0.767110661		3.178435956	\\
0.553879833		3.300622104	\\
0.375157115		3.393085907	\\
0		3.585622611	\\
0		3.585585456	\\
0		3.585538679	\\
0		3.585479787	\\
0		3.585405644	\\
0		3.585312298	\\
0		3.585194773	\\
0		3.585046804	\\
0		3.584860499	\\
0		3.584625919	\\
0		3.584330544	\\
0		3.583958599	\\
0		3.583490206	\\
0		3.582900308	\\
0		3.582157311	\\
0		3.581221367	\\
0		3.58004218	\\
0		0	\\
};
\addlegendentry{\scriptsize IB TG \cite{chaaban2016capacity}}

\addplot [color=black]
  table[row sep=crcr]{%
0	0\\
4.15725772303727	0\\
4.34503933573848	0\\
4.53279012457265	0\\
4.63582239098733	0\\
4.63792338498693	0\\
4.64616362913798	0\\
3.66726226452299	0.951698599406471\\
3.10254113475325	1.48781197844523\\
2.70918093273521	1.8529000526353\\
2.40957649230478	2.12466743686752\\
2.16766718918776	2.33883137281399\\
1.96811170406834	2.51095753230928\\
1.79728379819772	2.65420314413666\\
1.647166723849	2.77611647398313\\
1.40201642411468	2.96631911484071\\
1.19876729584063	3.11233982042397\\
1.0756782646172	3.19292837072485\\
0.893236678259252	3.30809366718204\\
0.739030900935834	3.38818334322114\\
0.473176432391389	3.51865793437814\\
0	3.69930434670581\\
0	3.69428232748172\\
0	3.69222031492763\\
0	3.68444779886742\\
0	3.67425830257574\\
0	3.66218124265805\\
0	0\\
};
\addlegendentry{\scriptsize IB ESDU \eqref{eq:bc_rate}}


\addplot [color=red]
  table[row sep=crcr]{%
0	0\\
4.1271031730895	0\\
4.24851676867426	0\\
4.42915119296486	0\\
4.61656550026481	0\\
4.62007408702805	0\\
4.62662741225169	0\\
4.63777810584697	0\\
4.63816111890594	0\\
1.91348916447904	2.44291644713767\\
1.73542029450344	2.59738071620257\\
1.57896004734873	2.72722993719825\\
1.3197583052718	2.92835027444795\\
1.10496100695082	3.07873101796348\\
0.956981972791032	3.16845403851442\\
0.776042268223243	3.27708194238774\\
0.496100557148972	3.43353300278806\\
0	3.68270753941175\\
0	3.67876795297022\\
0	3.6674668790459\\
0	3.66456634559892\\
0	3.66229903391011\\
0	3.65621227977071\\
0	3.64337167527226\\
0	3.63663510144145\\
0	0\\
};
\addlegendentry{\scriptsize IB ESDU \eqref{eq:bc_rate_esdu}}

\addplot [color=white,draw=none,  mark size=2.0pt, mark=x, mark options={solid, red}]
table[row sep=crcr]{%
0	3.6674668790459\\
0.610046011452507	3.3691263544581\\
1.13289537279867	3.03164555227827\\
1.52143802931413	2.74217933789597\\
1.8566122216989	2.45381705451177\\
2.08038349870072	2.26251574037814\\
2.32606482527153	2.02816843865516\\
2.35607510502945	2.02630968276133\\
2.64825836560249	1.75370275858687\\
2.66748962016755	1.74011465986047\\
2.67472822440109	1.7290439351557\\
3.05482734981722	1.40274341345466\\
3.06925949395574	1.39465123375737\\
3.07586236985236	1.38771898334701\\
3.07855711673658	1.38171515060336\\
3.07943534865768	1.37646556634054\\
3.07951553064718	1.37183685792714\\
3.63137075509387	0.884837957649042\\
3.64107472345745	0.881079313646723\\
3.64668955786514	0.877687667845631\\
3.64971402700313	0.874612408873138\\
3.65116412674885	0.871811647243173\\
3.65169476567981	0.869250473343887\\
3.6517097717274	0.866899592645087\\
3.65144840999863	0.864734258144286\\
3.65104738723156	0.862733431703816\\
3.65058225824676	0.860879121127354\\
3.65009351876161	0.859155852579927\\
3.64960224825816	0.857550247800218\\
3.64911908621685	0.856050682963812\\
3.64864918885459	0.854647011593952\\
3.64819487821261	0.853330338053171\\
3.64775702189362	0.852092831248543\\
3.64733573974115	0.850927570517032\\
4.61656550026481	0\\
};
\addlegendentry{\scriptsize IB \eqref{eq:bc_rate_esdu} with $\Delta_0=3\sigma_1$}

\end{axis}

\end{tikzpicture}%
\label{fig:bcb}}
	
	\subfloat[$\frac{A}{\sigma_1}=20$dB, $\sigma_2=10\sigma_1$]{\input{fig_snr20_sig2_10}\label{fig:bcc}}\hspace{1cm}
	\subfloat[$\frac{A}{\sigma_1}=30$dB, $\sigma_2=10\sigma_1$]{\input{fig_snr30_sig2_10}\label{fig:bcd}}
	\caption{Capacity inner and outer bounds for the peak-constrained two-user Gaussian BC.}
	\label{fig:bc}
\end{figure*}

Fig. \ref{fig:bc} shows the inner bounds (IB)  in \eqref{eq:bc_rate}, its computable form \eqref{eq:bc_rate_esdu} stated in Theorem \ref{thm:rate}, a benchmark inner bound from \cite{chaaban2016capacity} based on a TG distribution, and the outer bound (OB) stated in Theorem \ref{lem:bc_outer}, under various values of $\frac{A}{\sigma_1}$ and $\frac{\sigma_2}{\sigma_1}$.
To plot the rate region \eqref{eq:bc_rate_esdu}, we vary $\Delta_0$ within $\{0.5i\sigma_1|i=1,\dots,20\}$. Then, for each $\Delta_0$, we let $K=\max\bigl\{2,\lceil\frac{A}{\Delta_0}\rceil+1\bigr\}$ and $K_1$ in $\{0,\ldots,K\}$. For each $K_1$, we choose $K_2$ to be the smallest integer such that $\Delta=\frac{A}{K_1K_2-1}\le\Delta_0$. The same procedure is used for evaluating \eqref{eq:bc_rate}, except that $\Delta_0$ \cc{here is within $\{i\sigma_1|i=1,\dots,10\}$ for the sake of less computation time}. 
It can be seen from Fig. \ref{fig:bc} that the ESDU-based inner bound \eqref{eq:bc_rate} always outperforms the TG-based one, and the gap between the inner bound \eqref{eq:bc_rate} and its simplified form \eqref{eq:bc_rate_esdu} is within 0.2 bits in all tested cases. \cc{The gap is mainly attributed to the relatively loose ESDU lower bound as shown in Fig. \ref{fig:esdu_rate}.}  
\cc{As an observation, it is also worth to note that the settings of $K$ and $K_1$ given in this simulation can help to achieve the rate pairs close to the boundary of the IB \eqref{eq:bc_rate_esdu} under each $\Delta_0$, which is more significant around the maximum sum-rate point. This can be seen from the rate pairs associated with each $\Delta_0$, where the case $\Delta_0=3\sigma_1$ is provided as an example in Fig. \ref{fig:bc}. }

\section{Conclusions and Open Questions}
We studied the achievable rate region of an evenly-spaced discrete distribution (ESDU) in a peak-constraint Gaussian broadcast channel (BC). To this end, we derived new lower and upper bounds for the ESDU rate achieved in a Gaussian channel, i.e., $I(X;X+Z)$ with $X$ following a ESDU distribution and a Gaussian noise $Z$. We provided numerical results to examine the analytical results. The ESDU-based BC inner bound is shown to outperform the benchmark inner bound in the literature, which is based on a truncated Guassian (TG) distribution. Besides, the obtained ESDU rate upper bound for the P2P channel is remarkably tight in all tested settings.

Future work can target tightening the lower bound on ESDU rate, which can help to close the gap between the approximation and the actual ESDU inner bound. Moreover, the work can be extended to consider non-uniform distributions over an evenly-spaced alphabet (such as geometric distribution \cite{farid2009channel,farid2010capacity}) which is useful for peak- and average-constrained channels that model Li-Fi applications. 


\appendices
\section{Proof of Lemma \ref{lem:cu}} \label{app:lem_cu}
Firstly, the lower bound $I(X;X+Z)\ge \underline{C}(A,\sigma)$ follows from \cite[Thm. 5]{Lapidoth-Moser-Wigger-2009}. 
$\overline{C}(A,\sigma)$ is the combination of the capacity upper bounds of the $[0,A]$-peak-constrained Gaussian channel in \cite[Thm. 5]{Lapidoth-Moser-Wigger-2009} and \cite[(12)]{thangaraj2017capacity}, so that it is direct to obtain $I(X;X+Z)\le \overline{C}(A,\sigma)$
It remains to prove the upper bound $I(X;X+Z)\le\frac{1}{2}\log\Bigl(1+\frac{A^2}{12\sigma^2}\Bigr)$. We have
\begin{subequations}
	\begin{align}
		I(X;X+Z) &= h(X+Z) - h(Z) \\
		& \le \frac{1}{2}\log\bigl(2\pi e\mathbb{V}[X+Z]\bigr) - \frac{1}{2}\log\bigl(2\pi e\sigma^2\bigr)\nonumber \\
		& = \frac{1}{2}\log\Bigl(1+\frac{A^2}{12\sigma^2}\Bigr),
	\end{align}
\end{subequations}
where the inequality follows since the Gaussian distribution maximizes the differential entropy under a variance constraint, and the last equality follows since $\mathbb{V}[X+Z]=\mathbb{V}[X]+\mathbb{V}[Z]=\frac{A^2}{12}+\sigma^2$. This ends the proof.

\section{Proof of Lemma \ref{lem:lb_esdu}} \label{app:lem_lb_esdu}
We first prove that $I(X;X+Z)\ge\mathsf{F}_1(A,K,\sigma)$. Let $\hat{X}$ be the nearest-neighbor estimator of $X$ from $X+Z$. Then, using Fano's inequality, we have
\begin{subequations}
	\begin{align}
		I(X;X+Z)&=H(X)-H(X|X+Z) \\
		& \ge H(X) - H\bigl(\mathbb{P}(\hat{X}\ne X)\bigr) \notag\\ 
		&\qquad - \mathbb{P}(\hat{X}\ne X)\log(K-1).  \label{eq:lb_esdu_temp1}
	\end{align}
\end{subequations}
To calculate $\mathbb{P}(\hat{X}\ne X)$, let $\Delta=\frac{A}{K-1}$ and 
\begin{equation}
	p_0 = \mathbb{P}(\hat{X}\ne X|X=0)=\mathsf{Q}\Bigl(\frac{\Delta}{2\sigma}\Bigr).
\end{equation}
Note that $\mathbb{P}(\hat{X}\ne X|X=A)=p_0$, and for $x\in\mathcal{X}\setminus\{0,A\}$ we have $\mathbb{P}(\hat{X}\ne X|X=x)=2p_0$. Thus, 
\begin{subequations}
	\begin{align}
		\mathbb{P}(\hat{X}\ne X) &= \sum_{x\in\mathcal{X}}\mathbb{P}(\hat{X}\ne X|X=x)\mathbb{P}(X=x)  \\
		& = \frac{2(K-1)p_0}{K} = \xi_{A,K}. \label{eq:lb_esdu_temp2}
	\end{align}
\end{subequations}
By substituting \eqref{eq:lb_esdu_temp2} into \eqref{eq:lb_esdu_temp1}, we obtain $I(X;X+Z)\ge\mathsf{F}_1(A,K,\sigma)$.

Then, we prove that $I(X;X+Z)\ge\mathsf{F}_2(A,K,\sigma)$.  Define an independent random variable $U\sim{\rm Unif}\bigl([0,\Delta)\bigr)$. We have 
\begin{subequations}
	\begin{align}
		&\quad I(X;X+Z) \notag\\
		& = h(X+Z) - h(Z) \\
		& = h(X+U+Z|U) - h(Z) \\
		& = I(X+U;X+U+Z) - I(U;X+U+Z) \label{Eq:F2_c}\\
		&\ge I(X+U;X+U+Z) - I(U;U+Z) \\
		&\ge \underline{C}\Bigl(K\Delta,\sigma\Bigr) - \mathsf{E}(\Delta,\sigma),
	\end{align}
\end{subequations}
where the first inequality follows since $I(U;U+Z)-I(U;X+U+Z)=I(X;U|X+U+Z)\ge0$, and the second inequality follows since $X+U$ and $U$ are continuous uniform distribution where $X+U\sim{\rm Unif}([0,A+\Delta])$, so that Lemma \ref{lem:cu} applies. By substituting $\Delta=\frac{A}{K-1}$, we obtain $I(X;X+Z)\ge\mathsf{F}_2(A,K,\sigma)$. 

\cc{Finally, $I(X;X+Z)\ge\mathsf{F}_3(A,K,\sigma)$ is obtained by applying Jensen's inequality in $h(X+Z)$, as shown in \cite[eq. (18b)]{dytso2016interference}. }

\section{Proof of Lemma \ref{lem:ub_esdu}} \label{app:lem_ub_esdu}
The proof of $I(X;X+Z)\le \min\{H(X), \overline{C}(A,\sigma)\}$ is trivial. Next we prove that $I(X;X+Z)\le\mathsf{G}'(A,K,\sigma)$. Define $\Delta=\frac{A}{K-1}$ and define an independent random variable $U\sim{\rm Unif}\bigl([0,\Delta)\bigr)$. We start similar to \eqref{Eq:F2_c} to write
	\begin{align*}
		I(X;X+Z) = I(X+U;X+U+Z) - I(U;X+U+Z).
	\end{align*}
Then we continue as follows
\begin{subequations}
	\begin{align}
		&\quad I(X;X+Z) \\
		& = I(X+U;X+U+Z) - h(X+U+Z) + h(X+Z) \nonumber \\
		& \le I(X+U;X+U+Z) \\ 
		&\quad - \frac{1}{2}\log\bigl(2^{2h(U)}+2^{2h(X+Z)}\bigr) + h(X+Z)\nonumber \\
		& = I(X+U;X+U+Z) - \frac{1}{2}\log\Bigl(1+ \frac{2^{2h(U)}}{2^{2h(X+Z)}}\Bigr) \\
		& = I(X+U;X+U+Z) - \frac{1}{2}\log\Bigl(1+ \frac{2^{2h(U)}}{2\pi e \sigma^2 2^{2I(X;X+Z)}}\Bigr)\nonumber
	\end{align}
\end{subequations}
where the inequality follows by applying the entropy power inequality to lower bound $h(X+U+Z)$ with $\frac{1}{2}\log\left(2^{2h(U)}+2^{2h(X+Z)}\right)$. After some manipulations, this leads to
\begin{subequations}
	\begin{align}
		I(X;X+Z) &\le \frac{1}{2}\log\Bigl(2^{2I(X+U;X+U+Z)}-\frac{2^{2h(U)}}{2\pi e\sigma^2}\Bigr) \\
		&\le \frac{1}{2}\log\Bigl(2^{2\mathsf{E}(A+\Delta,\sigma)}-\frac{2^{2h(U)}}{2\pi e\sigma^2}\Bigr)
	\end{align}
\end{subequations}
where the second inequality follows since $X+U\sim{\rm Unif}([0,A+\Delta])$ so that $I(X+U;X+U+Z)\le\mathsf{E}(A+\Delta,\sigma)$ from Lemma \ref{lem:cu}. This ends the proof.

\bibliographystyle{IEEEtran}
\bibliography{ref_lib}

\end{document}